# Deep Learning in Software Engineering

Xiaochen Li[1], He Jiang[1,2], Zhilei Ren[1], Ge Li[3], Jingxuan Zhang[4]

[1]School of Software, Dalian University of Technology

[2]School of Computer Science & Technology, Beijing Institute of Technology

[3]Software Institute, Peking University

[4]College of Computer Science and Technology, Nanjing University of Aeronautics and Astronautics

li1989@mail.dlut.edu.cn, jianghe@dlut.edu.cn, zren@dlut.edu.cn,

lige@pku.edu.cn, jingxuanzhang@mail.dlut.edu.cn

## Abstract

Recent years, deep learning is increasingly prevalent in the field of Software Engineering (SE). However, many open issues still remain to be investigated. How do researchers integrate deep learning into SE problems? Which SE phases are facilitated by deep learning? Do practitioners benefit from deep learning? The answers help practitioners and researchers develop practical deep learning models for SE tasks. To answer these questions, we conduct a bibliography analysis on 98 research papers in SE that use deep learning techniques. We find that 41 SE tasks in all SE phases have been facilitated by deep learning integrated solutions. In which, 84.7% papers only use standard deep learning models and their variants to solve SE problems. The practicability becomes a concern in utilizing deep learning techniques. How to improve the effectiveness, efficiency, understandability, and testability of deep learning based solutions may attract more SE researchers in the future.

## Introduction

Driven by the success of deep learning in data mining and pattern recognition, recent years have witnessed an increasing trend for industrial practitioners and academic researchers to integrate deep learning into SE tasks [1]-[3]. For typical SE tasks, deep learning helps SE participators extract requirements from natural language text [1], generate source code [2], predict defects in software [3], etc. As an initial statistics of research papers in SE in this study, deep learning has achieved competitive performance against previous algorithms on about 40 SE tasks. There are at least 98 research papers published or accepted in 66 venues, integrating deep learning into SE tasks.

Despite the encouraging amount of papers and venues, there exists little overview analysis on deep learning in SE, e.g., the common way to integrate deep learning into SE, the SE phases facilitated by deep learning, the interests of SE practitioners on deep learning, etc. Understanding these questions is important. On the one hand, it helps practitioners and researchers get an overview understanding of deep learning in SE. On the other hand, practitioners and researchers can develop more practical deep learning models according to the analysis.

For this purpose, this study conducts a bibliography analysis on research papers in the field of SE that use deep learning techniques. In contrast to literature reviews, bibliography analysis can reflect the overview trends, techniques, topics on deep learning in SE by statistical data. First, we collect 4,443 research papers that contain both SE and deep learning related keywords. Next, we filter the research papers by reading their contents, citations and references. Finally, 98 research papers related to both SE and deep learning are identified. With these papers, we analyze the publication trend, research topics, used deep learning models, and industrial research interests of these papers.

We find that research papers related to deep learning has increased significantly in SE in recent years, which have facilitated 41 SE tasks. Both communities of SE and Artificial Intelligent (AI) show great interests in utilizing deep learning in SE. Surprisingly, more than one fifth research papers have industrial practitioners to participate in, which means that industrial practitioners are also interested in integrating deep learning into their SE solutions. Despite the encouraging success of deep learning, we find several concerns about using deep learning in SE. Practitioners and researchers worry about the practicability of utilizing a complex method with almost opaque internal representations like deep learning [6]. Hence, the effectiveness and efficiency [7], understandability [8], and testability [9] become the burden to use deep learning in practice. Fortunately, recent studies have conducted some initial investigation on these problems [6]-[9]. These findings may guide the future studies of using deep learning in SE.

## Example of using deep learning in SE

Deep learning is a technique that allows computational models composed of multiple processing layers to learn representations of data with multiple levels of abstraction [14]. In this section, we present an example of using deep learning in SE. In this example, we apply the deep learning model AutoEncoder on a typical SE task, i.e., bug reports summarization [10].

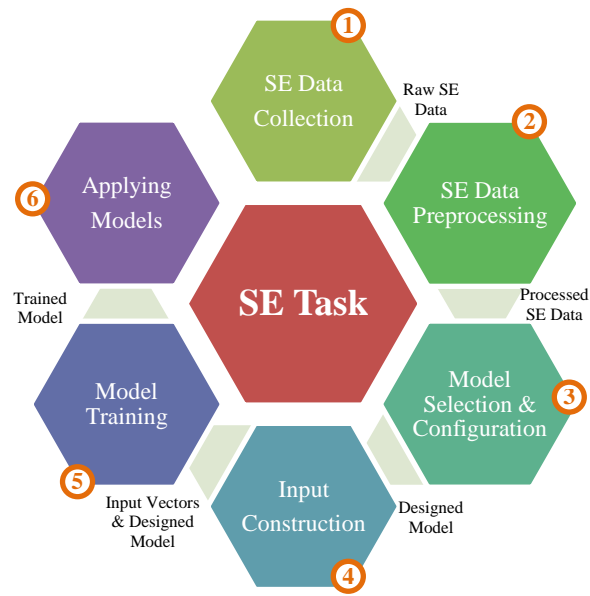

Fig. 1. A framework to summarize bug reports with AutoEncoder

Bug reports are texts to describe the bugs in software. Facing numerous bug reports, bug report summarization aims to generate a summary by extracting and highlighting informative sentences of a bug report to shorten the reading time. To identify informative sentences, researchers utilize AutoEncoder to encode the words in bug report sentences in an unsupervised way. Since the hidden state of AutoEncoder provides a compressed representation of the input data, the weights of words in a bug report can be measured by calculating how much information of a word is reserved in the hidden states. Based on the word weights, informative sentences are identified [10]. As shown in Fig. 1, the example consists of six steps.

1. **SE data collection** decides the available data for an SE task. For bug report summarization, the commonly available data are bug reports. Each bug report mainly contains a title, a description of the bug, and several comments.

2. **SE data preprocessing** removes the noises in SE data. For a bug report, the English stop words and some programming-specific ones are the noises. Besides, extremely short sentences are also noises, since they may be uninformative.

3. **Model selection and configuration** select suitable deep learning models for SE data and decide model configurations, e.g., the number of layers and neural units of each layer. The widely used deep learning models include AutoEncoder, CNN, RNN, etc. (explained in Fig. 3). These standard models usually have several variants, e.g., LSTM, Bi-LSTM, and attention-based RNN are classical variants of RNN. In this example, AutoEncoder is selected. AutoEncoder usually has a symmetric architecture, i.e., the number of neural units of input and output layers are the same. The output layer is defined as a pattern to reconstruct the input layer. The number of neural units of hidden layers decreases as towards the middle of the network. After training, the hidden states reserve the key information for reconstructing the input layer.

4. **Input construction** transforms SE data into vectors for deep learning models. For bug report summarization, researchers calculate the word frequency in bug reports and transform the word frequency values into vectors. These vectors are regarded as a training set for AutoEncoder.

5. **Model training** trains the designed model with the training set. A deep learning model usually has thousands of parameters representing the weights of connections among neural units. Hence, training the model is to tune these parameters according to the training set. For AutoEncoder, the parameters are trained by minimizing the difference between the input and output layers in an unsupervised way.

6. **Applying models** is to utilize the trained model to solve SE problems. In this example, the trained model can encode the word frequency vector of a new bug report into the hidden states. We can trace and calculate the changes of the value in each vector dimension along with the encoding process, and then deduce the weights of words in each dimension. These word weights help researchers assign weights of the sentences and select informative ones.

## Data collection

To collect deep learning related papers in SE, we design three criteria to search research papers from four well-known digital libraries, including Web of Science, ACM Digital Library, IEEE Xplore, and Scopus.

C1. Research papers should contain at least one of the following SE phrases, including "software engineer*", "software develop*", "software test*", "software design", "requir* analysis", "software requir*", "software maintain*", and "software manag*". The sign "*" is a wildcard character to match zero or more characters in a word.

C2. Research papers should contain at least one phrase about deep learning concept, i.e., "deep learn*" and "neural network*".

C3. Research papers are conference or journal papers written in English on the topic of computer science.

Under these criteria, we achieve 4,443 candidate research papers published before March 2018, including 414 from Web of Science, 207 from ACM Digital Library, 2,271 from IEEE Xplore, and 1,551 from Scopus. We remove the duplicate papers and short papers with less than 4 pages. At last, 3,351 research papers are reserved. We download and manually examine the contents of the papers:

1. We remove 35 papers that the full-contents cannot be downloaded.

2. We remove 2,441 papers that the searching phrases in C1 and C2 merely match some supplementary information in the paper. For example, "software engineer*" may match the phrase of "school of software engineering" in author biography or the publication venue "Transaction on Software Engineering".

3. After step 1 and 2, another 812 papers are removed as they do not focus on SE or deep learning. For example, "deep learning" is also a concept in computer education and "neural network" may refer to a shallow network structure with a single hidden layer.

At last, 63 research papers are remained. We take these papers as seeds to search their references and citations. If a new SE research paper about deep learning is found, we recursively examine the new paper. Finally, another 35 research papers are found. Hence, we collect in total 98 published or accepted research papers for analysis.

## Bibliography Analysis

We analyze the collected papers to investigate the status of deep learning in SE.

### A. The prevalence of deep learning in SE

We count the number of research papers each year and the venues of the publications in Fig. 2(a) and Fig. 2(b) respectively. In Fig. 2(a), we find that deep learning attracts little attention in SE for a long time, i.e., only less than 3 papers are published each year before 2015. The reason may be that although deep learning performs well on image processing, speech recognition, etc., it takes time for the practitioners and researchers in SE to validate deep learning on domain-specific SE tasks. However, the researches boom in SE after 2015, e.g., 28 publications in 2016 and 39 publications in 2017. Furthermore, only in the first three months in 2018, there are already 12 publications using deep learning in SE.

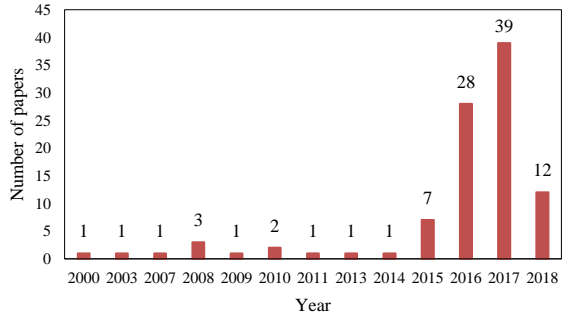

(a) The number of publications per year

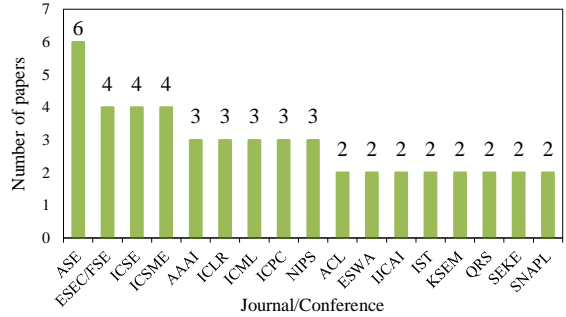

(b) Top venues of the publications

(c) Full names of publication venues

| Venue | Explanation |
|---|---|
| **ICSE** | Int'l Conf. on Softw. Eng. |
| **ESEC/FSE** | Joint European Softw. Eng. Conf. and Symposium on the Foundations of Softw. Eng. |
| **ASE** | Int'l Conf. on Automated Softw. Eng. |
| **ICSME** | Int'l Conf. on Softw. Maintenance and Evolution |
| **AAAI** | AAAI Conf. on Artificial Intelligence |
| **ICLR** | Int'l Conf. on Representation Learning |
| **ICML** | Int'l Conf. on Machine Learning |
| **ICPC** | Int'l Conf. on Program Comprehension |
| **NIPS** | Conf. on Neural Information Processing Systems |
| **ACL** | Annual Meeting of the Association for Computational Linguistics |
| **ESWA** | Expert Systems With Applications |
| **IJCAI** | Int'l Joint Conf. on Artificial Intelligence |
| **IST** | Information and Softw. Tech. |
| **KSEM** | Int'l Conf. on Knowledge Science, Eng. and Management |
| **QRS** | Int'l Conf. on Quality Softw |
| **SEKE** | Int'l Conf. on Softw. Eng. and Knowledge Eng. |
| **SNAPL** | Summit on Advances in Programming Languages |

Fig. 2. Basic information of deep learning in SE

For these research papers, we count the publication venues. Surprisingly, out of the 98 SE papers, 66 venues have published at least one paper on the topic of deep learning. Fig. 2(b) presents the publication venues that publish more than one paper We explain these venue names in Fig. 2(c). We find that using deep learning in SE attracts the attention from both communities of SE and AI, including some premier SE venues like ICSE, ESEC/FSE, ASE, ICSME, ICPC and some renowned AI venues like AAAI, ICLR, ICML, NIPS, ACL, IJCAI. These venues may be a good guidance to study the progress of deep learning in SE.

To conclude, deep learning is prevalent in SE. It attracts the attention from both SE and AI communities.

### B. The way to integrate deep learning into SE

As the prevalence of deep learning in SE, we analyze the way to integrate deep learning into SE. Fig. 3 shows the name of deep learning models and the number of papers using these models. We find that most studies (55 papers) directly transfer standard deep learning models into SE, including AutoEncoder, CNN, DBN, RNN and a simple fully-connected DNN. Meanwhile, the classical variants of these models in AI are also widely used (28 papers) such as SDAEs, LSTM, etc. The above models are used in 84.7% research papers. Besides using a single model, combined deep learning models (8 papers) also show competitiveness in SE, e.g., a combination of RNN and CNN. For the remaining papers, researchers design specific deep learning architectures for SE data like Stepped AutoEncoder and TBCNN. The above phenomenon suggests that when integrating deep learning into SE tasks, a new practitioner may be willing to first try some standard models and their variants to investigate whether deep learning works or not.

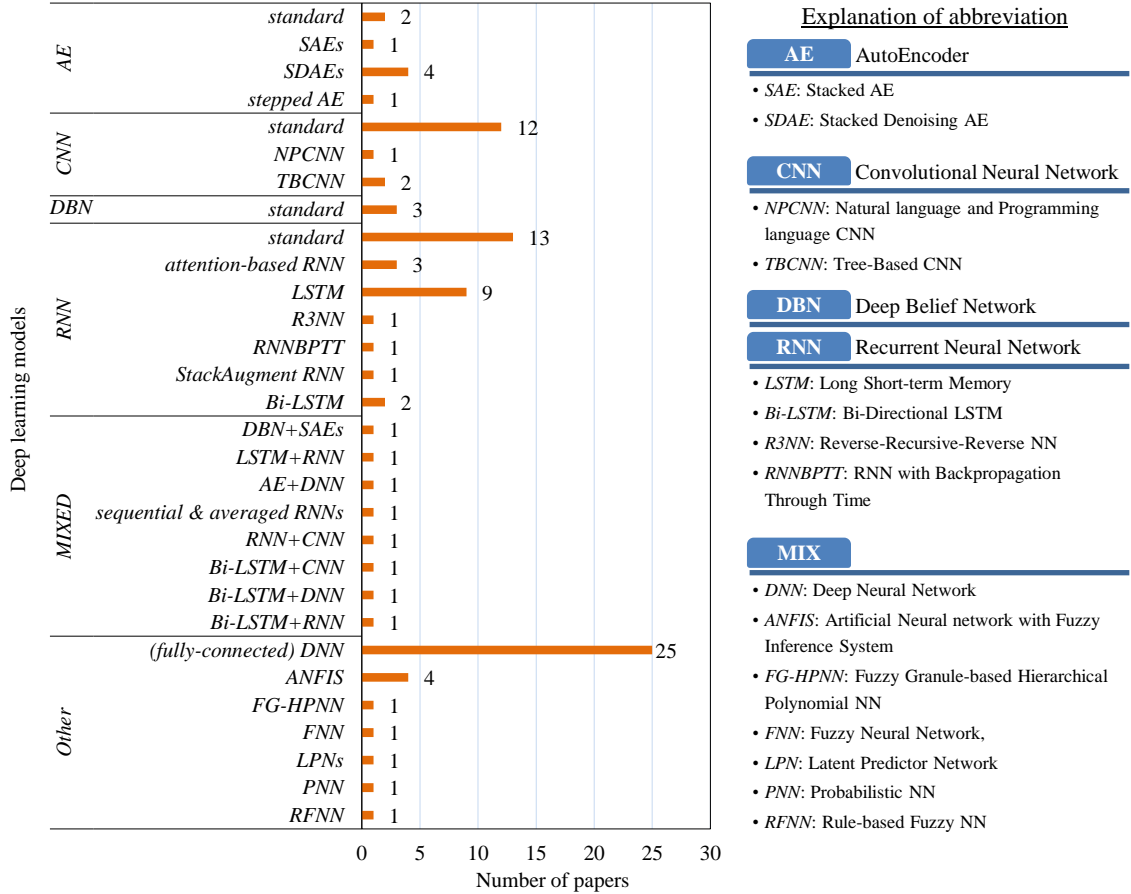

Fig. 3. Deep learning models in the research papers

Furthermore, we investigate what types of SE data are usually fed into these models. We analyze the inputs of the 98 papers. The inputs can be categorized into five categories.

1. Predefined software metrics (25 papers). Researchers first manually define and calculate some software metrics, e.g., lines of code, the number of bugs in source code. Then, they construct vectors based on the values of these metrics to feed into deep learning models.

2. Dynamic software status (14 papers). This category takes the dynamic information when running the software as input, e.g., the CPU utilization, the invoked APIs. The values of these dynamic information can be transformed into vectors for deep learning models.

3. Raw text or source code without sequence (32 papers). These papers treat the bag-of-words of text and source code as deep learning input without considering word sequences [10]. Based on the bag-of-words, word embedding, one-hot representation and word frequency are widely used to transform words into vector space for deep learning.

4. Raw text or source code in sequence (22 papers). In contrast to category 3, this category considers the sequence of words [2]. Such inputs are usually associated with RNN-based models, which utilizes the order of words to predict the next word or class label of software documents and source code, e.g., program learning and program synthesis.

5. Others (5 papers). Most of the other inputs are multimedia data such as images. For example, researchers utilize images to test deep learning models [9]. The pixels of the images are fed into deep learning models.

To conclude, practitioners and researchers can achieve competitive results on more than 80% SE problems when only using standard deep learning models and their variants. Deep learning can well handle many types of SE data, including predefined software metrics, dynamic software status, and raw text or source code.

**C. The SE phases facilitated by deep learning**

Due to the diversity of SE tasks, it is important to identify the existing SE tasks facilitated by deep learning, since it helps practitioners find the potential to leverage deep learning in their own problems.

As suggested by classical SE models, i.e., Waterfall Model and Incremental Model [11], SE can be divided into five phases, including requirement analysis, software design, development, testing and maintenance. In addition, since SE is an activity involving many stakeholders (developers, testers, project managers, etc.), we also add project management as an SE phase. Fig. 4 shows the SE tasks facilitated by deep learning in the six phases.

As shown in Fig. 4, researchers have tried deep learning on at least 41 SE tasks. In requirement analysis, deep learning helps requirement analysts automatically extract requirements from natural language texts [1]. In software design, design patterns of software can be recognized [12]. In software development, deep learning helps developers on 14 SE tasks from 30 research papers, including program learning and program synthesis [2], code suggestion [6], etc. Besides, software testing and maintenance are also major phases to attempt deep learning. There are 54 research papers in these two phases which cover 21 SE tasks like defect prediction [3] and reliability or changeability estimation [4]. For the 41 SE tasks, program learning and program synthesis [2], malware detection [5], defect prediction [3], reliability or changeability estimation [4], and development cost or effort estimation [13] are the top 5 tasks studied by the researchers. Hence, practitioners may have a board selection of methodologies and deep learning models when using deep learning on these tasks.

To conclude, deep learning has facilitated at least 41 SE tasks in all SE phases, including requirement analysis, software design, development, testing, maintenance, and project management.

Fig. 4. The SE tasks solved by deep learning and participated by industrial practitioners

## D. Research interests of industrial practitioners

We analyze the SE tasks participated by industrial practitioners to understand research interests in practice.

The industrial practitioners are identified when at least one author affiliation in the author list of a research paper is a company. In Fig. 4, we label the industry-participated SE tasks in bold and list the company names. To our surprise, there are 21 research papers (more than one fifth) on 13 SE tasks with at least one industrial practitioner, which implies the interest of industrial practitioners in integrating deep learning into SE problems. Among the 13 tasks, program learning and program synthesis attract the most attention [2]. Eight research papers from four companies have tried deep learning on this task, including DeepMind, Facebook, Google, and Microsoft. Besides, practitioners also apply deep learning on SE tasks like malware detection [5], development cost or effort estimation [13], etc., and achieve competitive results. Hence, deep learning may be useful on these tasks from the perspective of practitioners. This finding provides a guidance for academic researchers to apply deep learning in practice.

However, we find a mismatch from the top researched SE tasks and the industrial interests. For the top 5 tasks studied by deep learning in Fig. 4, only program learning and program synthesis, and malware detection attract more than one industrial practitioner to participate in. The reason may be that, on the one hand, practitioners have not found a suitable way to apply deep learning on other SE tasks in practice. On the other hand, practitioners already select some lightweight methods to solve these tasks. Hence, there is still a long way to apply a complex method like deep learning in industry.

To conclude, practitioners participate in more than one fifth research papers. They benefit from deep learning on 13 SE tasks, including program learning and program synthesis, malware detection, etc.

### E. Concerns to use deep learning in SE

Despite the prevalence of improving SE tasks with deep learning, many concerns emerge on the practicability of using deep learning in SE [6]. As a complex and almost opaque model, several factors limit the practicability of deep learning, including the effectiveness, efficiency, understandability, and testability. These issues may influence the development of deep learning in SE.

**Effectiveness and Efficiency**. Recent studies show that by applying a simple optimizer Differential Evolution to fine tune SVM, it achieves similar results with deep learning on linking the knowledge unit in Stack Overflow [7]. Most importantly, this method is 84 times faster than training deep learning models. The same phenomenon is also observed on code suggestion, in which an adapting n-gram language model specifically designed for software surpasses RNN and LSTM [6]. Although techniques like off-line training and cloud computing may partially resolve the efficiency problem [10], there is still a tradeoff between deep learning and other lightweight, domain-specific models. Such tradeoff drives a deep investigation on deep learning, e.g., what types of SE data and tasks are suitable for deep learning and how to integrate the domain knowledge into deep learning.

**Understandability**. The understandability is a burden to "control" deep learning. Recently, several methods are proposed to improve the understandability of deep learning. For example, practitioners in Facebook explore to visualize industry-scale deep neural networks [8]. The proposed tool help software engineers understand the neuron activations, individual instances, classification results, and differences between activation patterns of deep learning. Such tool is a good start to increase the understandability of deep learning in SE.

**Testability**. As a complex model, the testability limits the security of applying deep learning in SE. Hence, researchers attempt to use software testing techniques to improve the testability of deep learning, i.e., deep learning testing [9]. To test deep learning models, coverage testing and metamorphic testing are recently applied [9]. Coverage testing validates whether all the neural units in deep learning are correctly activated. Metamorphic testing generates the test oracle for coverage testing. These studies demonstrate the importance of SE techniques on validating artificial intelligence techniques like deep learning.

To conclude, the practicability of deep learning is still a rising and hot topic for SE practitioners and researchers.

## Conclusion

Deep learning recently plays an important role for solving SE tasks. In this study, we conduct a bibliography analysis on the status of deep learning in SE. We find that deep learning has been integrated into more than 40 SE tasks by both industrial practitioners and academic researchers. Most studies use standard deep learning models and their variants to solve SE problems. The practicability of deep learning may hider SE practitioners from using deep learning in practice, which is a rising and hot topic for further investigation.